# Archaeology in the Digital Age: From Paper to Databases

Frédérique Mélanie-Becquet*, Johan Ferguth*, Katherine Gruel** & Thierry Poibeau*

*Laboratoire LATTICE
CNRS & Ecole normale supérieure & Université Paris 3 Sorbonne Nouvelle

**Laboratoire AOROC
CNRS & Ecole normale supérieure & Ecole Pratique des Hautes Etudes

45 rue d'Ulm 75005 Paris

firstname.lastname@ens.fr

Research units in archaeology often manage large and precious archives containing various documents, including reports on fieldwork, scholarly studies and reference books. These archives are of course invaluable, recording decades of work, but are generally hard to consult and access. In this context, digitizing full text documents is not enough: information must be formalized, structured and easy to access thanks to friendly user interfaces.

The situation at AOROC, a research unit of Ecole normale supérieure specialized in archaeology, is precisely the one described hereabove: several decades of research are contained in documents, which are hardly accessible, even for people working in the lab. The situation is such that researchers may produce studies largely overlapping with previous work, which remained unknown because of its poor accessibility.

A partnership has thus been established between AOROC and LATTICE, another research unit specialized in Digital humanities and natural language processing, to digitize and structure a part of these documents. A pilot study concerned a collection of texts covering excavations related to the Gaul period (*Cartes Archéologiques de la Gaule*, [Provost, 1988-]), over an area encompassing a large part of modern France from the Iron Age to the Medieval period (800BC to 800AD). 128 volumes have been published so far: each volume corresponds to one French department (some departments are covered by several volumes). The pilot study concerned three of these volumes, along with other types of documents so as to ensure the genericity of the developed solution. The idea is of course not just to digitize and transfer documents online but also to extract key information so as to feed structured databases (Poibeau et al., 2013). The result should be accessible using a standard but powerful query language.

A first step consists in recognizing the structure of the documents, which mainly consist in notices, each notice corresponding to a specific "municipality" (the structure is not formally encoded in the source documents and may vary from one document to the other). Specific zones inside the notices have to be recognized (see **figure 1**): this can be done by specific scripts but also needs some manual cleaning. In our opinion, the most interesting part concerns the natural language processing techniques used for information extraction. These include:

– Named entity recognition (i.e. the recognition of proper names, location, dates, etc.)
– Technical term extraction (i.e. all archaeological terms)
– Entity linking (i.e. the recognition of the different variants of a same term or entity and its connection to the same type of object in the database).

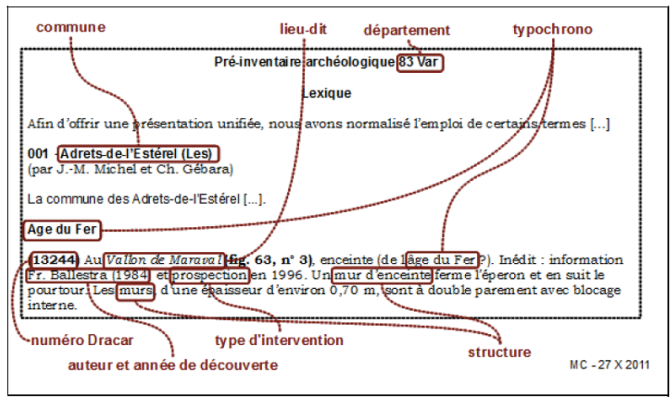

**Figure 1:** a typical notice with key information to be recognized

Different tools have been used, like TreeTagger (http://www.cis.uni-muenchen.de/~schmid/tools/TreeTagger/) for part-of-speech tagging, Yatea (https://metacpan.org/release/Lingua-YaTeA) for technical term extraction and TyDI for terminology structuration (http://ciam.inra.fr/logiciels/node/195). In-house solutions have been developed for document structure, named entity recognition and entity linking. All the modules are parameterized so that they can be easily adapted to new sources of data. These tools are based on up-to-date natural language processing techniques and have obtained excellent results in recent benchmarks like Semeval (Ruiz and Poibeau, 2015).

The result of the project made it possible to automatically feed a structured database based on the textual content analysis. Documents are now accessible online with full text facilities, structured indexes and ontologies (see **figure 2**). It is thus possible to interrogate the database with queries dealing with a specific location, a specific series of objects or a given period of time.

**Figure 2:** the original published text, along with some structured outputs after analysis

This work can be compared to other initiatives with a similar goal. In archaeology, one can cite archaeological archiving bodies such as the ADS (http://archaeologydataservice.ac.uk/), tDAR

([http://www.tdar.org/about/](http://www.tdar.org/about/)), or OpenContext ([http://opencontext.org/](http://opencontext.org/)), among others. Research on interoperability between archaeological databases includes, among many others (Binding et al., 2008; Doerr, 2003; Jordal et al., 2004; Vlachidis and Tudhope, 2012). Our project is different since it is from the beginning designed to deal with texts in different languages, especially French, German and English, with a cross-linguistic perspective. One of the major research issues is the maintenance of an international terminology referring to complex notions that can vary from one country to the other. The system should be flexible enough to be able to match related concepts (even if they vary slightly from one source to the other), but relevant enough so as to provide only relevant information. This goal involves a permanent dialogue between experts of the domain and the maintenance of an up-to-date terminology. The tools and interfaces developed within the project should help to keep this goal a reality as much as possible.

**Acknowledgements**

This work has received support of Paris Sciences et Lettres (program "Investissements d'avenir" ANR-10-IDEX-0001-02 PSL*) and of the laboratoire d'excellence TransferS (ANR-10-LABX-0099). The project has been realised during the PEPS CNRS-PSL EITAB 2013-2014.